# SimplySQS: An Automated and Reproducible Workflow for Special Quasirandom Structure Generation with ATAT

Miroslav Lebeda[a,b,c], Jan Drahokoupil[a], Petr Vlčák[a], Šimon Svoboda[a], Axel van de Walle[d]

[a] *Faculty of Mechanical Engineering, Czech Technical University in Prague, Technická 4, 16607 Prague 6, Czech Republic*

[b] *Faculty of Nuclear Sciences and Physical Engineering, Czech Technical University in Prague, Trojanova 339/13, 12000 Prague 2, Czech Republic*

[c] *FZU – Institute of Physics of the Czech Academy of Sciences, Na Slovance 2, 18200 Prague 8, Czech Republic*

[d] *School of Engineering, Brown University, 182 Hope Street, 02912 Providence, USA*

## Abstract

The special quasirandom structure (SQS) method is widely used for modeling disordered materials under periodic boundary conditions, with the ATAT *mcsqs* module being one of the most established implementations. However, SQS generation with *mcsqs* typically relies on manual preparation of input files, ad hoc execution scripts, and post-processing steps, which introduces user-dependent errors and limits reproducibility. Here, we present *SimplySQS* (https://simplysqs.com), an automated and reproducible workflow for SQS generation that is delivered through an online, interactive interface. *SimplySQS* guides users through structure import, compositional and supercell definition, and cluster parameter selection, while automatically generating all required ATAT input files and a single all-in-one execution script that encapsulates the complete search process. By standardizing input preparation, execution, and output analysis, the framework minimizes errors associated with manual file handling and enables consistent reproducibility of SQS searches. The workflow is demonstrated on the $Pb_{1-x}Sr_xTiO_3$ (PSTO, including $PbTiO_3$ (PTO) and $SrTiO_3$ (STO)) perovskite system. SQSs spanning the entire concentration range were generated using a single automated bash script produced by *SimplySQS*, after which all resulting structures were subjected to geometry optimization using a universal machine-learning interatomic potential (MACE MATPES-r$^2$SCAN-0). This approach reliably reproduced the experimentally observed cubic-to-tetragonal transition near $x \approx 0.5$, with lattice parameters deviating by less than 1 % in the cubic region ($x > 0.5$) and less than 4 % in the tetragonal region ($x \leq 0.5$). Overall, *SimplySQS* transforms SQS generation with ATAT into an intuitive, reproducible, and systematic framework for modeling disordered materials.

## 1. Introduction

Accurate and efficient modeling of disordered alloys is essential in materials science, as many technologically significant materials adopt disordered structures. One simple and transparent approach is to construct a large supercell and randomly assign elements to atomic sites. However, due to the finite size of the supercell, this method often results in local atomic arrangements that can deviate significantly from the ideal random state. The special quasirandom structure (SQS) approach [1] offers a systematic improvement over the random assignments. An SQS is a periodic supercell specifically designed so that its local atomic correlation functions closely approximate those of an ideal random alloy. Within the cluster expansion formalism, atomic configurations are described by symmetry-averaged correlation functions of clusters of lattice sites (pairs, triplets, etc.), whose target values for a fully random solution follow from independent site occupations. The SQS is then obtained by optimizing the atomic arrangement to minimize deviations from these random correlations, typically prioritizing short-range clusters. In doing so, SQS provides an efficient and reasonably accurate model for disordered systems under periodic boundary conditions. Over the years, SQSs have been widely employed in density functional theory (DFT) and molecular dynamics (MD) studies to investigate properties of various materials [2–10].

Several powerful software applications have been developed for generating SQSs. Among the most widely adopted are (1) the *mcsqs* package [11] in the Alloy Theoretic Automated Toolkit (ATAT) [12], (2) the Integrated Cluster Expansion Toolkit (ICET) [13], and (3) the Sqsgenerator Python package (github.com/dgehringer/sqsgenerator) [14]. ATAT, implemented in C++, is a long-established console-based tool, widely used in the community. ICET, a more recent Python-based framework, combines a Python interface with a C++ back-end, enabling efficient SQS generation and integration with common materials science packages such as pymatgen [15] and Atomic Simulation Environment (ASE) [16]. The Sqsgenerator package, built with a Python interface and a C++ back-end, provides fast parallel execution via OpenMP and optional MPI. Its main current limitation is that it generates SQSs for a single user-specified supercell at a time (with lattice vectors restricted to simple multiples of the unit cell) and that it is limited to pair and triplet clusters. Among these three tools, we seek to provide an automated and reproducible interactive workflow for the ATAT *mcsqs* due to its large established user base, the extent of its capabilities and its integration with the broader ATAT toolkit.

The ATAT *mcsqs* has significantly advanced the capabilities of researchers working in the field of disordered alloys. However, despite its power, the learning curve associated with it can pose a challenge, particularly for users without a strong programming background or experience with command-line environments. Preparing input structures, defining chemical configurations for atomic sites, specifying supercell parameters or number of atoms, and interpreting or converting output files require a deep understanding of the format settings and certain coding familiarity. However, as visible in the community forum (matsci.org/c/atat/), users often encounter various problems, mostly related to the input files. Many of these problems originate from misunderstandings of the required file formats. Therefore, standardizing and automating the preparation of input files as well as analysis of output files would greatly improve accessibility by reducing user-dependent errors, and it would enable systematic, reproducible SQS searches across different computational studies.

To address this, we have developed *SimplySQS* (simplysqs.com), an online, interactive browser-based workflow for the SQS generation with *mcsqs*. An advantage of the online platform is that it requires no compilation and can be used on any device with an internet connection and a web browser. For cases where online access may be unavailable, the application can also be compiled and run locally (see *Section 2: Implementation Details*). The primary goal of *SimplySQS* is to formalize SQS generation with *mcsqs* into a reproducible, automated workflow that standardizes input preparation, execution control, and output analysis. In doing so, *SimplySQS* lowers the entry barrier for newcomers while providing convenience for more experienced users. In addition, *SimplySQS* includes a lightweight assessment of random structures, allowing users to quickly evaluate whether a given supercell and a defined composition could already provide a statistically adequate representation of a random alloy, or whether SQS optimization is likely to offer a meaningful improvement.

In this manuscript, we describe the design and implementation of *SimplySQS* and demonstrate its capabilities through two examples: the preparation of input files for creating SQS supercell of $Sr_{0.6}Ba_{0.4}Ti_{0.6}Zr_{0.4}O_3$ perovskite, and the generation of a complete series of SQSs for $Pb_{1-x}Sr_xTiO_3$ in order to evaluate the evolution of lattice parameters and the energetics as a function of Sr content using the MACE MATPES-r$^2$SCAN-0 [17,18] universal machine-learning interatomic potential (MLIP). Finally, we discuss the potential impact of *SimplySQS* on improving accessibility, reproducibility, and educational adoption of the ATAT *mcsqs* for modeling disordered alloys.

## 2. Implementation Details

The *SimplySQS* is implemented in Python and deployed online (simplysqs.com) as an interactive web interface using the Streamlit framework (streamlit.io). It is currently hosted on the free Streamlit community cloud server, which provides direct deployment from the public GitHub repository (github.com/bracerino/atat-sqs-gui). The repository documentation also includes a tutorial for local installation.

Within the application, pymatgen [15] and ASE [16] are employed for structure import, manipulation, and file format conversion, while Matminer [19] is used to calculate partial radial distribution functions (PRDF). Numerical operations and tabular data management are handled by NumPy [20] and Pandas [21]. Structure visualization is provided through py3Dmol [22], and the graphs are created with Plotly and Matplotlib [23].

The crystal structures from the Crystallography Open Database (COD) [24] and the Materials Project (MP) [25,26] databases are obtained through their publicly available application programming interfaces (APIs). For MP, data retrieval is carried out using its corresponding Python package that wrap its APIs, whereas access to COD is facilitated via a Representational State Transfer (RESTful) API. Access to the Materials Cloud Three-Dimensional Structure Database (MC3D) [27] is made via Optimade API [28]. An overview of all Python packages grouped by their primary role used in *SimplySQS* is presented in **Tab. 1**.

| | |
|---|---|
| Web Interface and deployment | Streamlit |
| Structure handling | pymatgen, ASE, mp-api, Optimade, RESTful API |
| Data handling | NumPy, Pandas |
| Structure and data visualization | py3DMol, Plotly, Matplotlib |

**Tab. 1**: Python packages used in the implementation of *SimplySQS*, grouped by their functionality.

## 3. Application Workflow and Features

*SimplySQS* is designed to guide users through the entire workflow of preparing input files, running the SQS search via the ATAT *mcsqs* with a comprehensive all-in-one bash script, and analyzing output files (see the general workflow in **Fig. 1**). Currently, the main features of the application include:

- **Structure import**: Upload structures in common formats (POSCAR (VASP structure input file [29,30]), CIF (Crystallographic Information File), LMP (LAMMPS data format [31]), XYZ structure format with lattice) or retrieve them directly from databases such as MP, COD, and MC3D.
- **Configurable SQS setup**: Guided forms allow to define supercell sizes (number of atoms), elements and concentrations for atomic sites, cluster cutoff distances, the number of parallel SQS searches, and setting the total search time.
- **Automated ATAT input and script generation**: The platform creates all required basic *mcsqs* files ('*rndstr.in*', '*sqscell.out*') and a comprehensive bash script ('*monitor.sh*') that automates the *mcsqs* run while providing information about the objective function and calculation time directly into the console (otherwise not previously available with mcsqs) as well as automatic conversion of the *mcsqs* specific SQS output format into commonly used POSCAR format.
- **Randomness score benchmarking vs. supercell size:** For a set target concentration, the platform computes and compares the normalized pair-wise randomness score across supercell sizes (using ensembles of randomly generated configurations), helping users decide whether an SQS optimization is likely to provide an improvement or whether a sufficiently large random supercell is expected to be adequate representation.
- **Output log parsing and visualization**: The platform allows to upload ATAT *mcsqs* output files and automatically extract from them the convergence of the objective function, correlation values, or calculating partial radial distribution functions from the best SQS.
- **Direct export of created SQS**: Generated SQSs can be converted into widely used formats (POSCAR for VASP, CIF for e.g. CASTEP [32], LMP for LAMMPS, XYZ for e.g. CP2K [33,34]) for direct use in DFT or MD simulations.

The workflow of *SimplySQS* platform can be divided into four main stages: (i) structure input, (ii) SQS setup, (iii) input file generation, and (iv) analysis of ATAT outputs. A video tutorial demonstrating its use is also available at: implant.fs.cvut.cz/atat-sqs-gui. Below, we illustrate the main application functionality using the example of generating an SQS for the perovskite $(Sr_{0.6}Ba_{0.4})(Ti0.6Zr_{0.4})O_3$, starting from the parent $SrTiO_3$ structure.

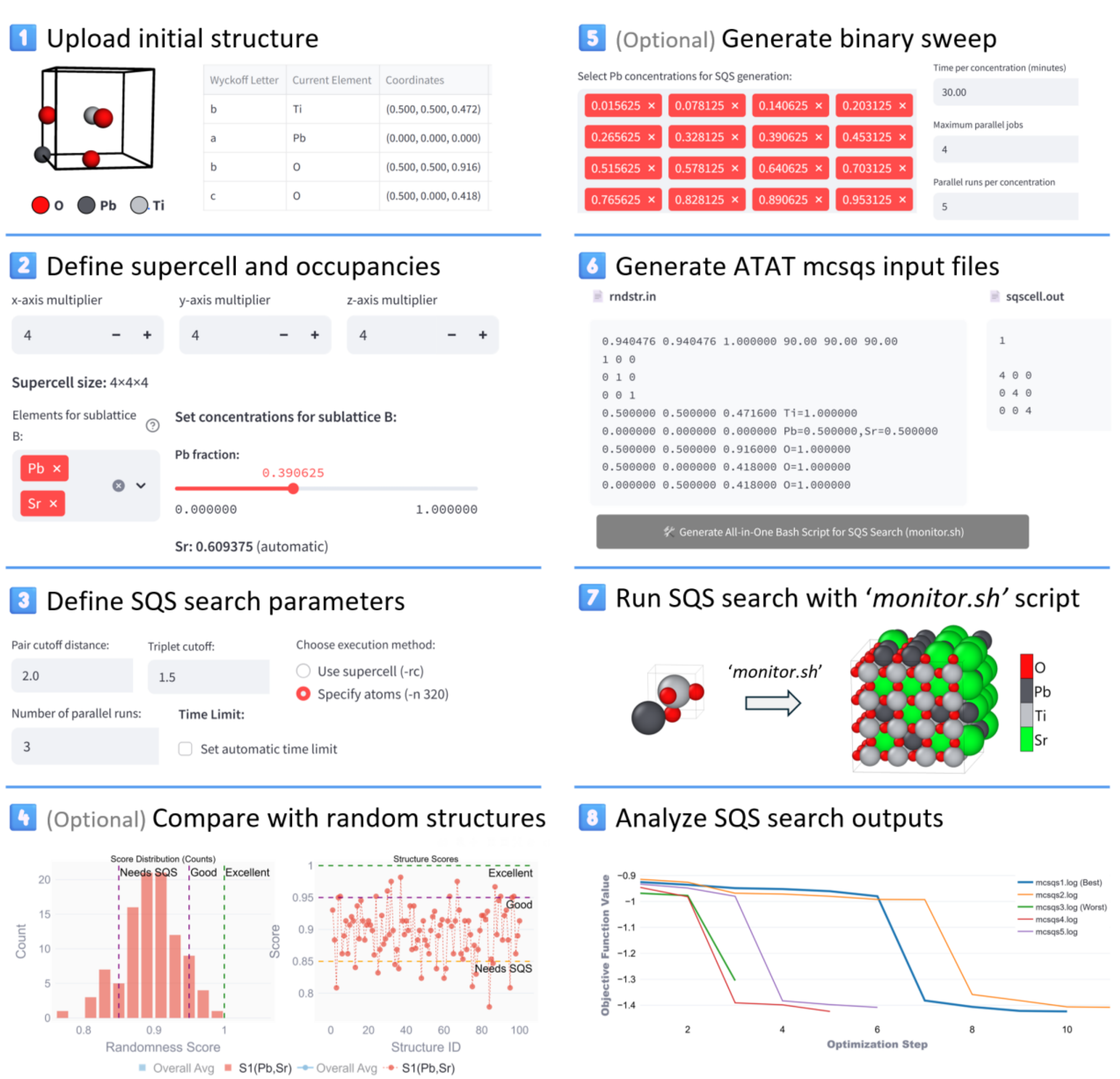

**Fig. 1:** Workflow illustrating the main features of *SimplySQS* for the reproducible generation of SQS with the ATAT *mcsqs* module.

### 3.1. Structure Input

The workflow begins with uploading the initial crystal structure that will define the initial atomic positions for the SQS generation. *SimplySQS* accepts several widely used structure file formats, including POSCAR (VASP), CIF, LMP (LAMMPS), and extended XYZ (with lattice information). In addition to direct file upload, the interface provides an integrated search tool for accessing structures from three crystal structure databases: MP, COD, and MC3D. Users can query these databases directly within the application and import structures without leaving the interface.

As an example, the initial structure of $SrTiO_3$ uploaded into the *SimplySQS* interface, is shown in **Fig. 2.** The application automatically identifies unique Wyckoff positions, which provide the sublattices basis for defining element configurations on them. In our example case, the Wyckoff indices corresponding to Sr and Ti define two sublattices, where Sr will be partially substituted with Ba, while Ti will be partially substituted with Zr to create SQS for $(Sr_{0.6}Ba_{0.4})(Ti_{0.6}Zr_{0.4})O_3$ perovskite. Moreover, the application offers an option to automatically convert the structure into its primitive cell.

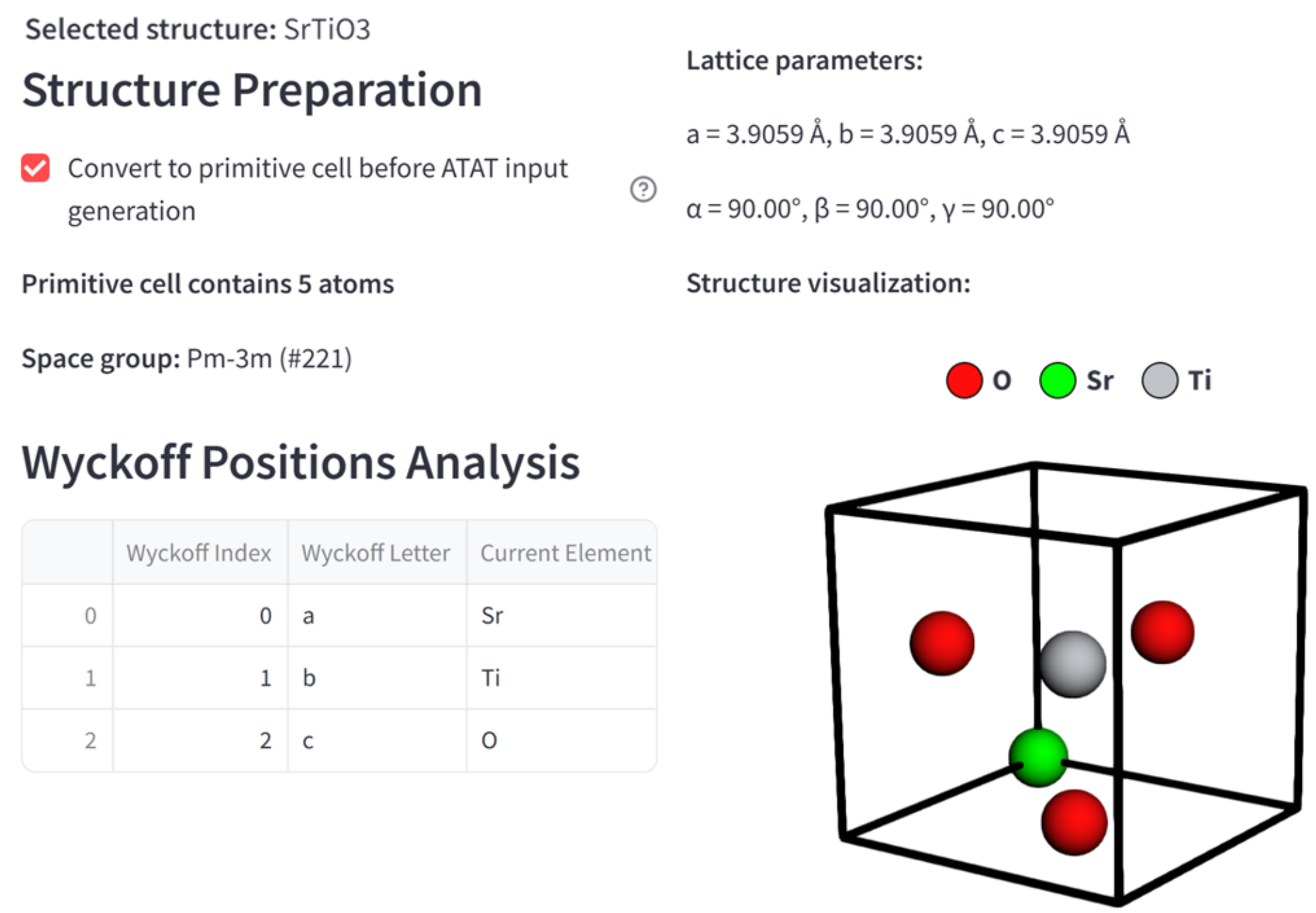

| | Wyckoff Index | Wyckoff Letter | Current Element |
|---|---|---|---|
| 0 | 0 | a | Sr |
| 1 | 1 | b | Ti |
| 2 | 2 | c | O |

**Fig. 2:** Uploaded initial $SrTiO_3$ perovskite crystal structure within the *SimplySQS* interface.

The workflow naturally supports multicomponent substitutional disorder, including ternary and higher-order alloys. In practice, such systems can be constructed by starting from an ideal bcc, fcc, or hcp parent lattice and assigning multiple elements with the desired concentrations. Low-dimensional materials can also be modeled when represented within periodic boundary conditions, for example by generating the SQS on a fully periodic lattice and subsequently introducing vacuum to form 2D slabs or layered supercells for atomistic simulations. Because SQS generation itself assumes periodicity in all directions, cluster cutoffs and the added vacuum thickness should be chosen carefully to minimize spurious interactions between periodic images along the non-periodic direction.

## 3.2. SQS Setup and Input File Generation

Once the base structure is imported, users can configure the target disordered alloy through guided forms. Two modes are available:

- **Global concentration mode**: specify overall species concentrations that apply to all atomic sites.
- **Sublattice mode**: define compositions for each crystallographic sublattice.

Supercell dimensions can be set either by specifying the exact size or for the SQS search, by constraining the number of atoms without fixing supercell shape. Once the supercell (number of atoms) is set, the available concentrations for each sublattice are automatically calculated within the interface. For the perovskite example, the sublattice mode was chosen, with 3 × 3 × 3 supercell of $SrTiO_3$, containing 135 atoms. This allows the creation of Ba and Zr concentrations close to 40 % - $(Sr_{0.59}Ba_{0.41})(Ti_{0.59}Zr_{0.41})O_3$. **Fig. 3** illustrates the interface for the configuration of sublattice A, corresponding to the initial element of Sr.

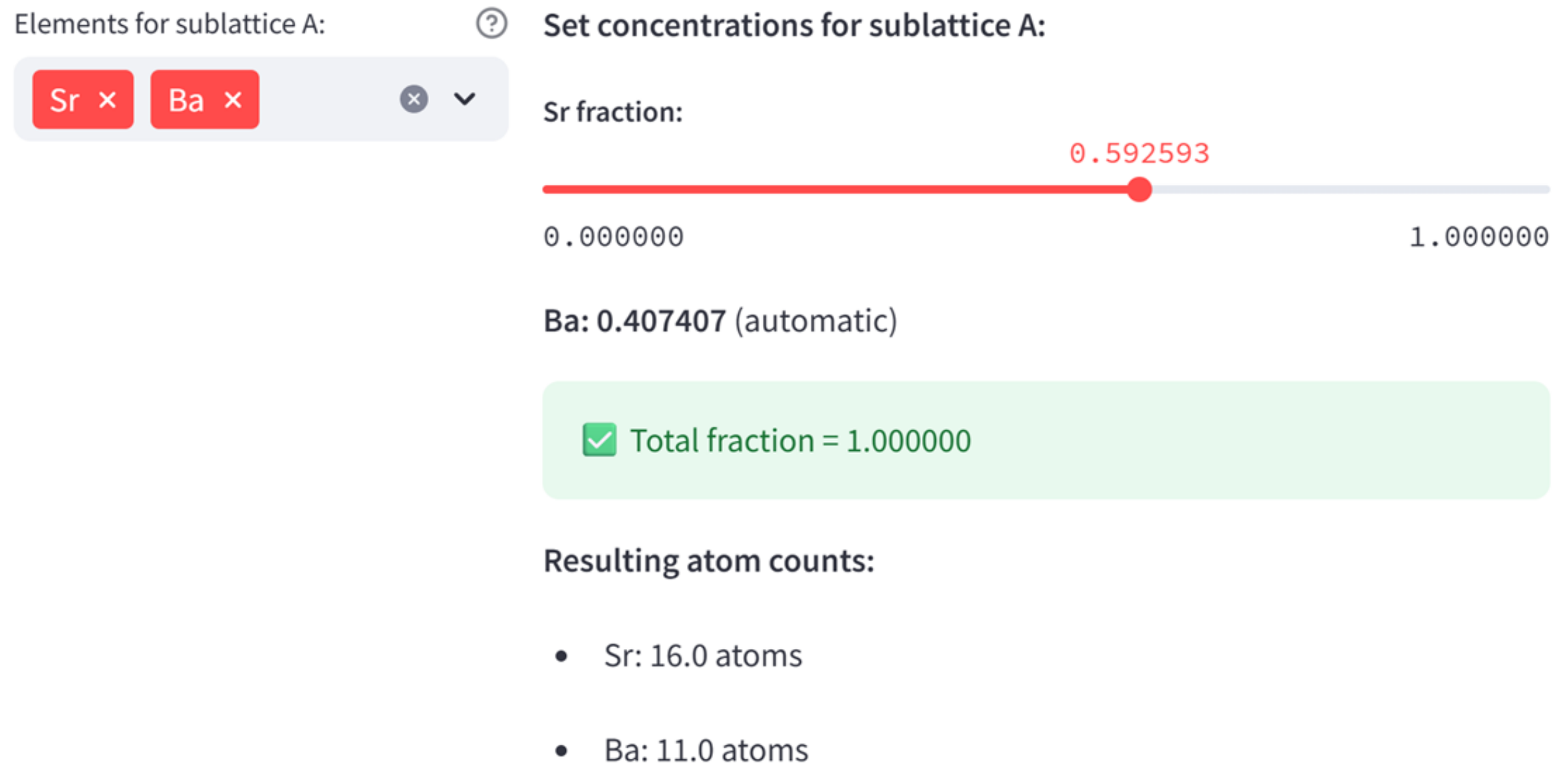

**Fig. 3:** Defining element types and concentrations for the A-site sublattice within the *SimplySQS* interface.

Furthermore, users can choose whether they wish to run the *mcsqs* in single or parallel mode. In parallel mode, independent searches for SQSs are launched with different initial seeds. For example, specifying 5 parallel runs will generate 5 SQS candidates. Additionally, users can adjust the cut-off distances for cluster interaction parameters (for pairs and optionally also for triplets and quadruplets). To simplify this step, *SimplySQS* can calculate nearest neighbor distances (up to the 6th shell, normalized to the largest lattice parameter) for the chosen structure and provides a histogram showing the number of pairs as a function of distance. For instance, as indicated in **Fig. 4,** including pairs up to the 4th nearest neighbor in the Sr-Ba sublattices corresponds to a pair cutoff of ~2.0 Å (normalized unit cell).

**Active sites:** 1/5 (Ba+Sr positions)

**NN Distances Between Active Sites (normalized to max lattice parameter):**

**1st NN:** 1.0000 | **2nd NN:** 1.4142 | **3rd NN:** 1.7321 | **4th NN:** 2.0000 | **5th NN:** 2.2361 | **6th NN:** 2.4495

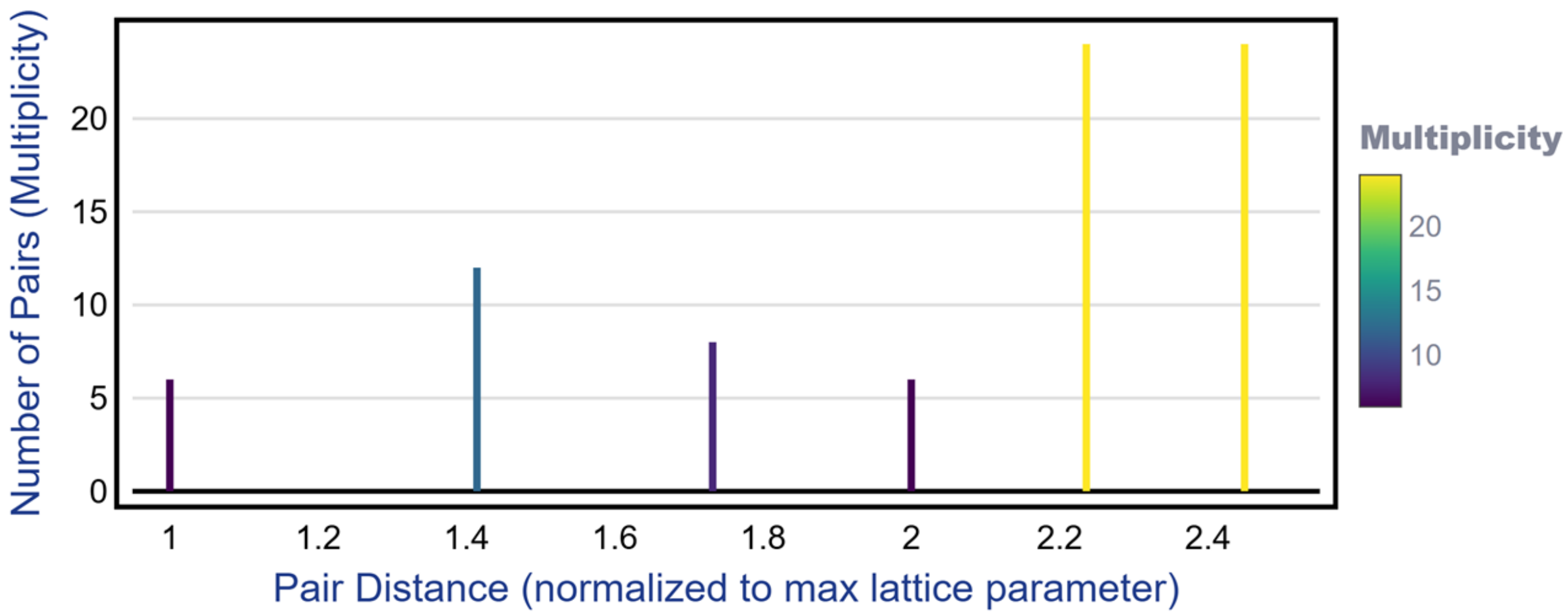

**Fig. 4:** Illustration of calculated nearest neighbor distances and the pair-distance histogram within the Sr-Ba sublattice in the *SimplySQS* interface.

Once the parameters are set, the platform automatically generates required ATAT input files (‘*rndstr.in’*, ‘*sqscell.out’*) along with an all-in-one bash script (‘*monitor.sh*’). This script completely streamlines the SQS search execution by:

1. Automatically creating input files with the specified parameters.
2. Starting the *mcsqs* search with the chosen cluster settings.
3. Printing real-time progress to the console, including elapsed runtime and best objective function values as well as creating a new file which contains time-dependent progress of objective function (‘*mcsqs_progress.csv’* for a single run or *‘mcsqs_parallel_progress.csv’* for a parallel run).
4. Allowing users to set a maximum runtime for the *mcsqs* search and, upon completion or manual termination, automatically converting the final ‘*bestsqs.out*’ SQS format into the POSCAR format, ensuring direct compatibility with VASP and other simulation packages.

An additional advantage of the all-in-one script is that it directly facilitates reproducibility: the same script can be easily shared and rerun on different computers, ensuring that the SQS search process can be reliably reproduced by other researchers.

The content of ‘*rndstr.in*’, ‘*sqscell.out’*, and the first command which generates the information about clusters for the set example of $(Sr_{0.59}Ba_{0.41})(Ti_{0.59}Zr_{0.41})O_3$ is shown in **Fig. 5**.

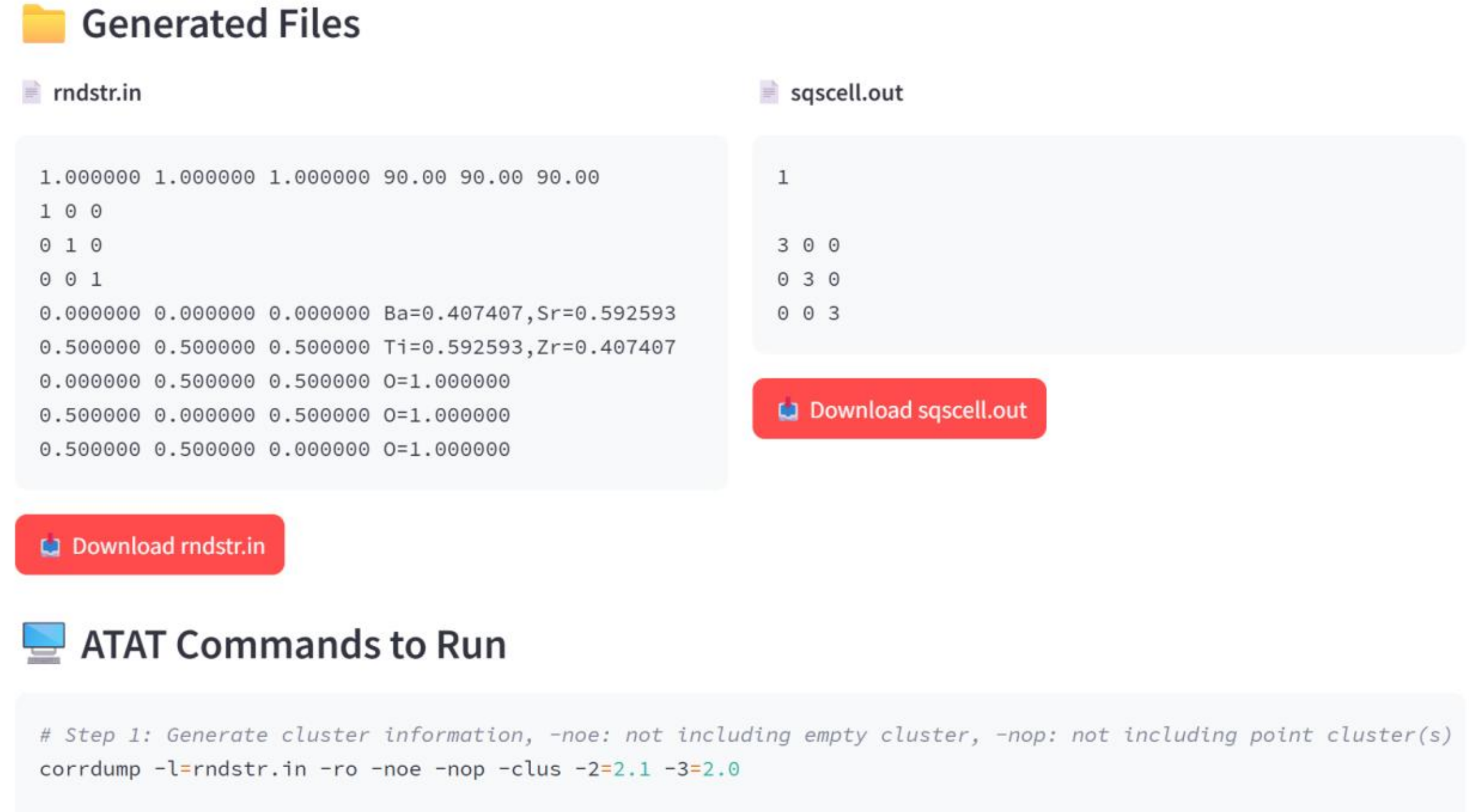

**Fig. 5:** Content of '*rndstr.in*', '*sqscell.out*' input files, and the first command to run that generates the cluster information for ATAT *mcsqs* search for SQS of $(Sr_{0.59}Ba_{0.41})(Ti_{0.59}Zr_{0.41})O_3$.

To further enhance usability, *SimplySQS* includes a new binary concentration sweep mode (an addition to the standard features of *mcsqs*). When users wish to generate SQSs across the concentration range of two elements (e.g., $Sr_{1-x}Ba_xTiO_3$, where $0 \leq x \leq 1$ ), the program automatically detects that the binary option has been selected and **the** user can use it to generate dedicated all-in-one bash script. This script automatically performs SQS searches for each concentration, creating a separate folder for every case and running the search for the user-defined total runtime (see **Fig. 6** for interface parameters). Once a search is complete, the best structure ('*bestsqs.out*') is automatically converted into POSCAR format, eliminating the need for manual conversion from mcsqs to formats compatible with external simulation packages. Users can also customize parallelization settings: (i) the number of independent *mcsqs* runs per concentration and (ii) the number of concentrations processed simultaneously. A video tutorial demonstrating this feature is available also at: implant.fs.cvut.cz/atat-sqs-gui.

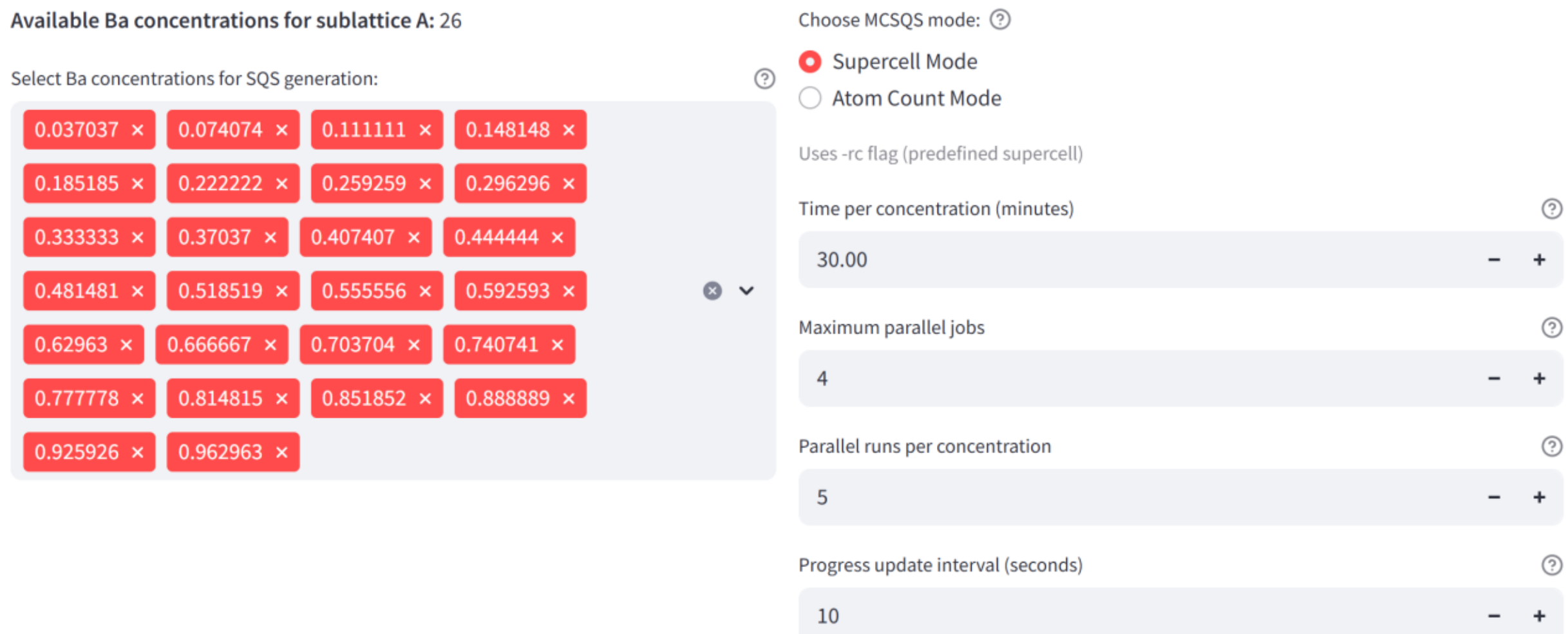

**Fig. 6:** Interface of the binary concentration sweep mode showing the configurable parameters: selected *mcsqs* mode, total search time, number of parallel runs per concentration, number of concentrations executed simultaneously, and the update interval for printing progress information to the console.

### 3.3 Analysis of ATAT *mcsqs* Outputs

The *SimplySQS* interface also facilitates post-processing and evaluation of results generated by ATAT *mcsqs*. In addition to simplifying the preparation of input files, it provides tools for converting output structures into standard formats and for analyzing the quality of the generated SQS.

Optimized SQS ('*bestsqs.out*') can be imported into the interface and converted into widely used formats (**Fig. 7**), including POSCAR, LMP, CIF, and extended XYZ (with lattice information). The best multiple SQSs obtained from parallel run can be also uploaded and batch-converted simultaneously. Moreover, PRDF from the best uploaded SQS can be directly calculated and visualized within the interface.

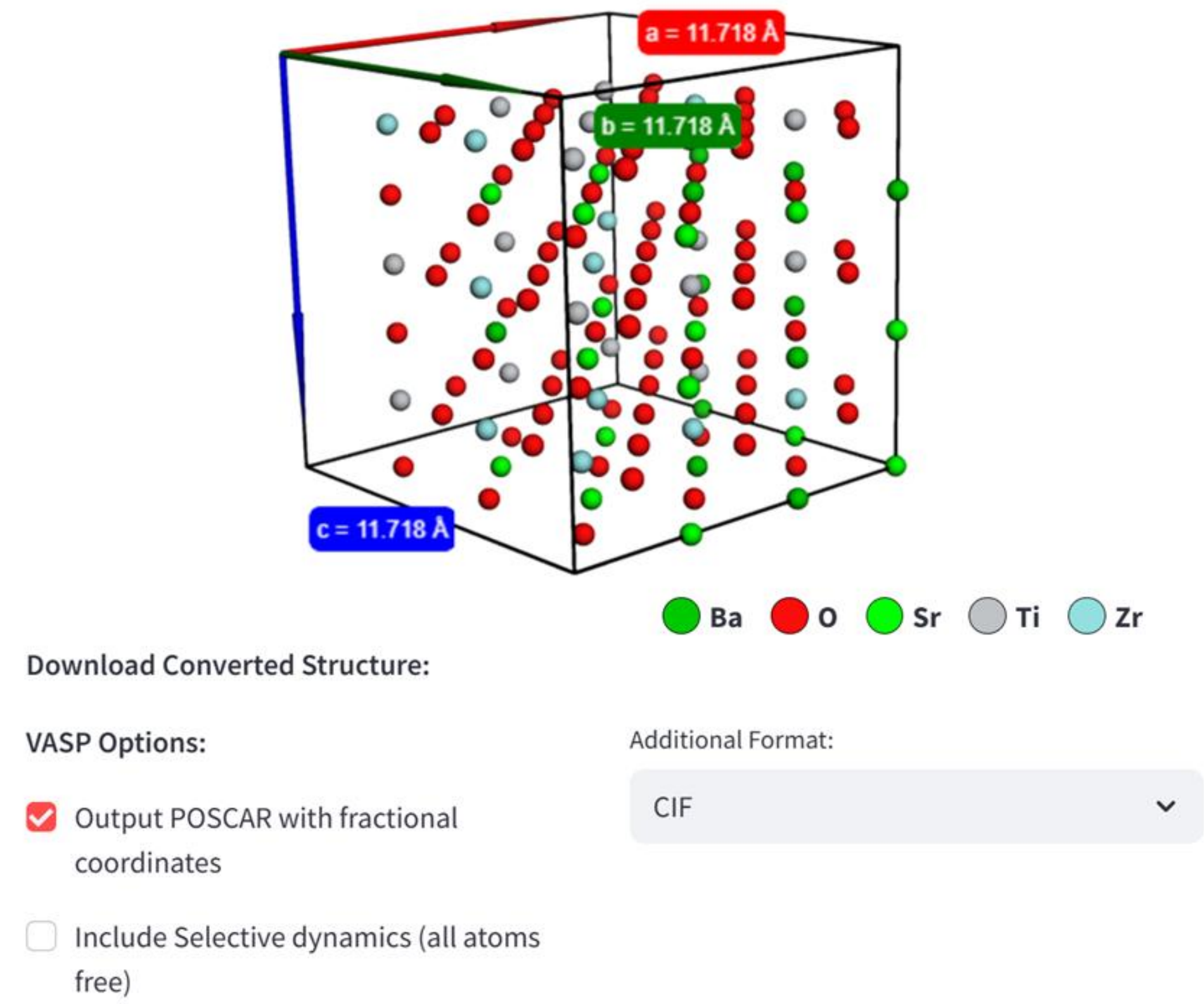

**Fig. 7:** Illustration of the best SQS for the $(Sr_{0.59}Ba_{0.41})(Ti_{0.59}Zr_{0.41})O_3$ uploaded into the *SimplySQS* interface, showing options to download the structure in different formats.

The platform also supports monitoring of the *mcsqs* search process. By uploading the main log file ('*mcsqs.log'*) or the progress files generated by the all-in-one bash script ('*mcsqs_progress.csv*' for single runs and '*mcsqs_parallel_progress.csv*' for parallel executions), users can visualize the convergence of the objective function with respect to search steps or elapsed time (**Fig. 8**) and immediately see which run yielded the best and which yielded the worst objective function. Moreover, *SimplySQS* can process the '*bestcorr.out*' file, which reports the cluster correlations of the final SQS. These are compared against the target values of an ideal random alloy, enabling users to evaluate how closely the generated SQS reproduces the desired random alloy.

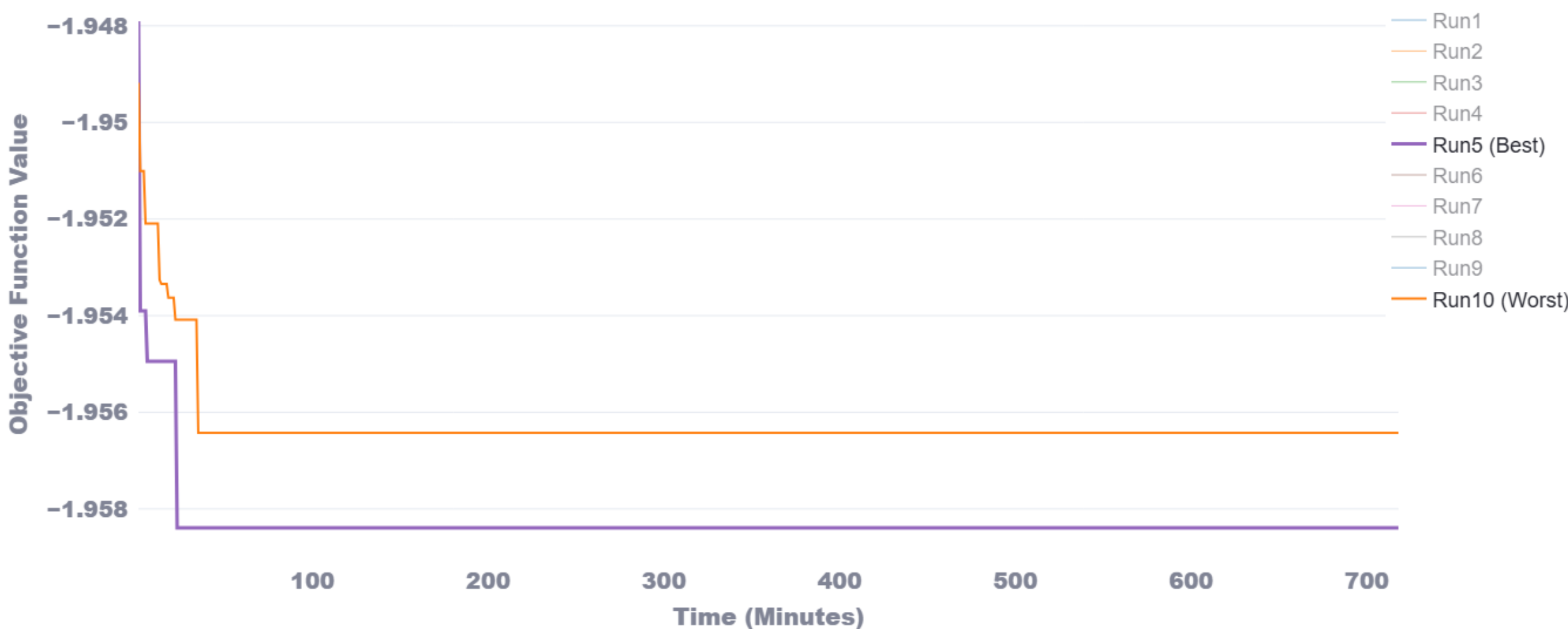

**Fig. 8:** Time-dependent convergence of the SQS search for $(Sr_{0.59}Ba_{0.41})(Ti_{0.59}Zr_{0.41})O_3$ using the ATAT *mcsqs* parallel run, highlighting the runs with the best and worst objective functions.

### 3.4 Random Structures versus SQS Quality Assessment

In addition to the automated SQS generation workflow based on ATAT *mcsqs*, *SimplySQS* includes a random-structure quality assessment module that allows users to determine *a priori* whether a given supercell and set of sublattice compositions require SQS optimization, or whether random atomic occupation already provides a statistically adequate representation of an ideal random alloy. This functionality addresses a common practical question in simulations of disordered materials, namely whether the computational cost of an SQS search is justified for a particular system size and composition, or whether randomly generated structures could be sufficient.

The assessment is based on a quantitative analysis of short-range order using Warren-Cowley (WC) parameters [35,36], evaluated independently for each chemically active sublattice and for the first three coordination shells under periodic boundary conditions. The analysis is limited to the pair correlations, which enable efficient and robust evaluation of short-range order in finite supercells. Pair correlations capture the earliest and strongest signatures of chemical ordering in most practical alloy systems, and significant non-random behaviour in higher-order correlations is typically already reflected at the pair level. For the purpose of rapidly assessing whether a given supercell and concentration set are sufficient, pairwise analysis therefore provides a balanced compromise between physical relevance and computational cost. When higher-order effects are expected to be important for a specific application, SQS optimization is recommended to be used.

Rather than relying on individual correlation values, *SimplySQS* introduces a normalized randomness score (from 0 to 1) that provides a compact measure of short-range order. For a given supercell and composition, multiple random configurations with the set supercell size and elements on the sublattices are generated independently. For each configuration, WC parameters are computed for all element pairs on each chemically active sublattice and for the first three coordination shells. The deviations of these parameters from ideal random alloy are first combined at the coordination-shell level using a root-mean-square measure over all species pairs. The resulting shell-wise deviations are then averaged over the first three coordination shells for each chemically active sublattice, and finally averaged over all active sublattices to obtain a single scalar measure of short-range order.

The overall randomness score is defined as one minus the averaged root-mean-square deviation of the WC parameters, such that a perfectly random structure yields a score of 1, while increasing short-range order results in progressively lower values. Based on the resulting score distribution, systems are classified as (i) sufficiently random (score ≥ 0.95), (ii) marginal, where SQS optimization will likely improve accuracy (0.85 ≤ score < 0.95), or (iii) significantly ordered, where the SQS search is strongly recommended (score < 0.85).

An example of the randomness score evaluation for 100 randomly generated 320-atom $Pb_{0.3}Sr_{0.7}TiO_3$ structures is shown in **Fig. 9**, with the supercell effect on the score within range of 3 × 3 × 3 (135 atoms) up to 8 × 8 × 8 (2560 atoms) shown in **Fig. 10**. The mean score for supercell 4 × 4 × 4 (320 atoms) is close to 0.9, indicating possible short-range order and suggesting that SQS optimization will probably produce structures that better approximate an ideal random alloy. In contrast, the larger 7 × 7 × 7 supercell (1715 atoms) exhibits an average score exceeding 0.95, approaching the ideal random limit. This implies that, for such large cells, simple random generation is likely sufficient to achieve adequate configurational randomness, making the SQS search rather unnecessary.

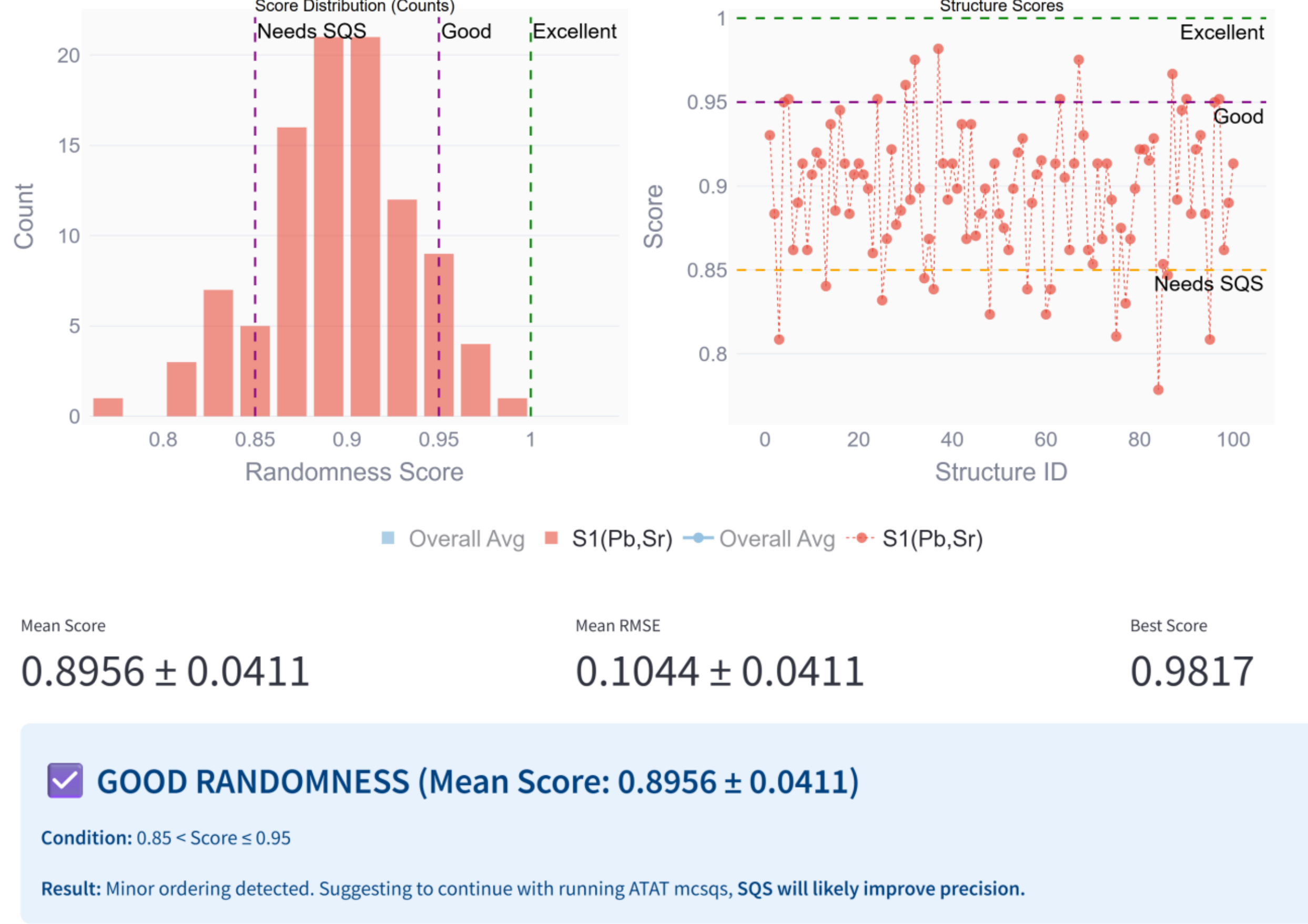

**Fig. 9:** Distribution of normalized randomness scores for randomly generated 320-atom $Pb_{0.3}Sr_{0.7}TiO_3$ structures (100 structures), as provided by the *SimplySQS* interface.

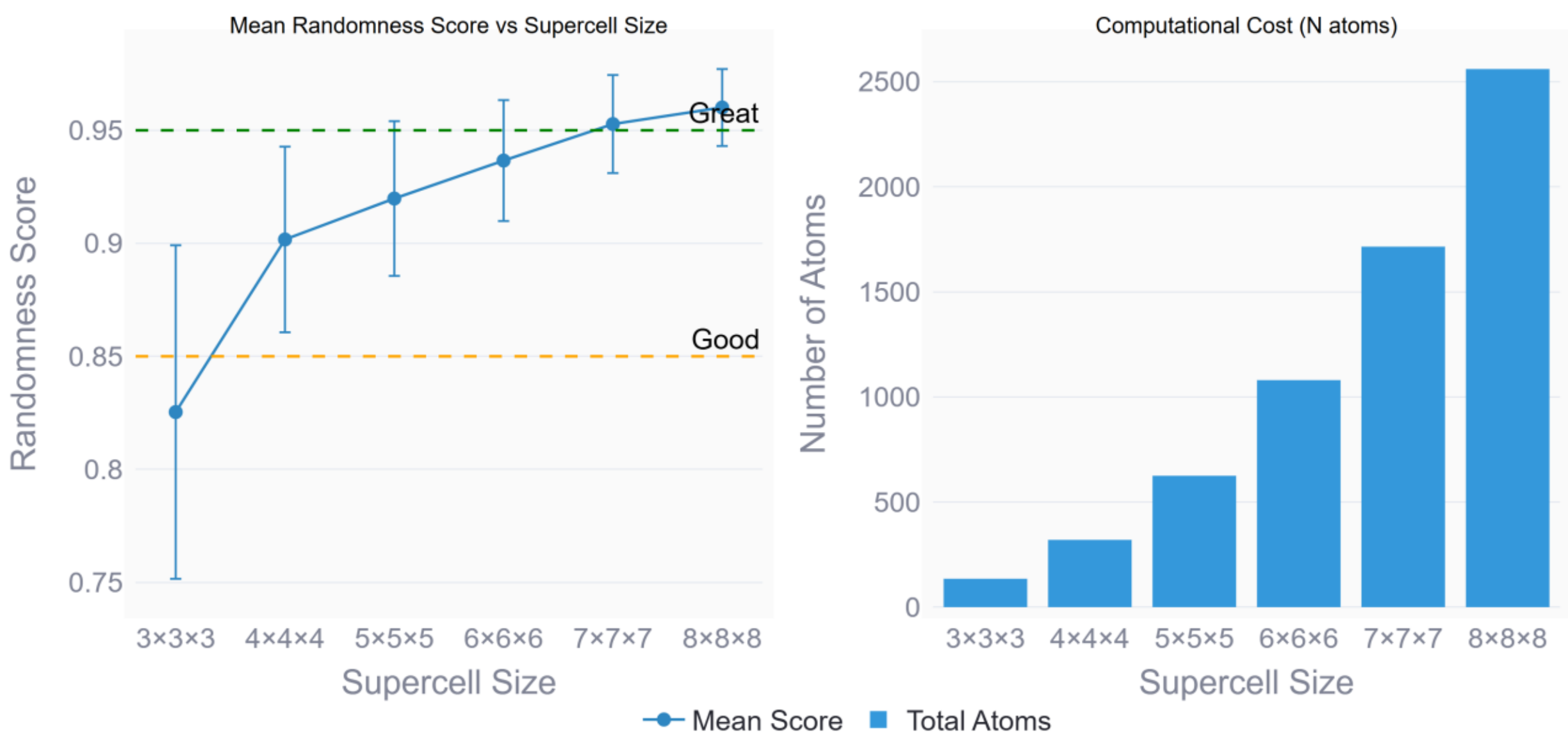

**Fig. 10: Left:** Variation of the randomness scores as a function of supercell size, each with 100 randomly generated supercells of $Pb_{0.29}Sr_{0.71}TiO_3$. **Right:** Number of atoms in the supercells as a function of supercell size. These plots are generated within the SimplySQS interface.

## 4. Example – $Pb_{1-x}Sr_xTiO_3$ Lattice Parameters and Mixing Energies with MACE r$^2$SCAN Universal MLIP

### 4.1 SQS Settings

To illustrate the effectiveness of *SimplySQS*, a series of SQSs was generated for the perovskite alloy system $Pb_{1-x}Sr_xTiO_3$ (PSTO, including $PbTiO_3$ (PTO) and $SrTiO_3$ (STO)) covering the full composition range. Each SQS comprised a 320-atom supercell, constructed as a 4 × 4 × 4 expansion of the initial $PbTiO_3$ unit cell. The SQS generation was performed using a single all-in-one bash script created via the binary concentration sweep mode of *SimplySQS*. The used script and generated SQS files are available on GitHub at: github.com/bracerino/atat-sqs-gui/tree/main/Example. Each concentration was searched for 4 hours, with five independent parallel runs per composition and 5 concentrations run in parallel. A pair-cluster cut-off distance of 1.5 Å and a triplet-cluster cut-off of 1.2 Å (in normalized unit cell units) were used. Notably, for all concentrations, the objective function exhibited no change in the last three hours of the search.

## 4.2 Geometry Optimization Settings and Mixing Energy

For each concentration, all five generated SQSs were employed for geometry optimization to obtain statistical changes in lattice parameters. Optimizations were performed with the universal (foundation) MACE MATPES-r2SCAN-0 MLIP. This foundation model was used because it was trained on MatPES (Materials Potential Energy Surface) DFT dataset [37] computed with the $r^2$SCAN (restored regularized Strongly Constrained and Appropriately Normed) exchange-correlation functional [38,39], which generally yields more accurate lattice parameters and energetics than conventional GGA-PBE (generalized gradient approximation Perdew-Burke-Ernzerhof) [40]. Here, its predictions are directly compared with reference experimental lattice parameters of the PSTO series, thus also assessing the model's quantitative accuracy for this system.

In each geometry optimization, both atomic positions and lattice vectors were relaxed while fixing cell angles and enforcing $a = b$ to avoid minor unintended deviations between $a$ and $b$ arising from finite supercell size effects. The starting lattice parameters for each concentration were taken from the previously optimized lower-Pb concentration, i.e., the sequence began with $SrTiO_3$ (initialized at the experimental cubic lattice constant of 3.905 Å [41]), and each subsequent composition used the optimized lattice of the prior step as its initial guess. A strict force convergence criterion of 0.002 eV · $Å^{-1}$ was applied, and the BFGS (Broyden-Fletcher-Goldfarb-Shanno) algorithm was used for the optimization in ASE. All optimizations converged in no more than 300 steps.

The mixing energies were evaluated from the relaxed total energies of the five SQSs (using their average total energy). All reported energies were normalized per atom (320 atoms in the supercells). The mixing energy was referenced to the end-member compounds according to

$$\Delta E_{\text{mix}}(x) = E_{\text{tot}}(\text{Pb}_{1-x}\text{Sr}_x\text{TiO}_3) - (1-x)E_{\text{tot}}(\text{PbTiO}_3) - xE_{\text{tot}}(\text{SrTiO}_3),$$

where $E_{\text{tot}}$ denotes the total energy per atom. Finite-temperature effects were approximated by including the configurational entropy associated with random occupation of the Pb/Sr sublattice (64 atoms). The ideal configurational entropy per atom was calculated as:

$$S_{\text{conf}}(x) = -k_B \frac{64}{320} [(1-x)\ln(1-x) + x\ln x].$$

The corresponding mixing free energy was then evaluated as

$$\Delta G_{\text{mix}}(x,T) = \Delta E_{\text{mix}}(x) - TS_{\text{conf}}(x).$$

Electronic and vibrational free-energy contributions were not included within the MLIP framework, such that the temperature dependence arises solely from configurational disorder.

## 4.3 Results

The optimized lattice parameters (**Fig. 11**) show a clear cubic-to-tetragonal phase transformation around x ≈ 0.5, consistent with experimental observations [42]. For $x < 0.5$, all structures remained cubic ($a = c$), and the deviation between simulated and experimental lattice parameters was under 0.3 %, indicating high precision of the MACE r$^2$SCAN potential in this region. For $x > 0.5$, a tetragonal distortion emerged: the simulated $a$ parameter decreases (and becomes lower than experiment, max deviation of 0.9 %) whereas the $c$ parameter increases (and becomes slightly higher than experiment, maximum deviation of 3.4 %) with increasing Pb content. We attribute this opposite trend to the increasing contribution of Pb-O bonding, which may not be fully captured by the universal potential's training DFT set.

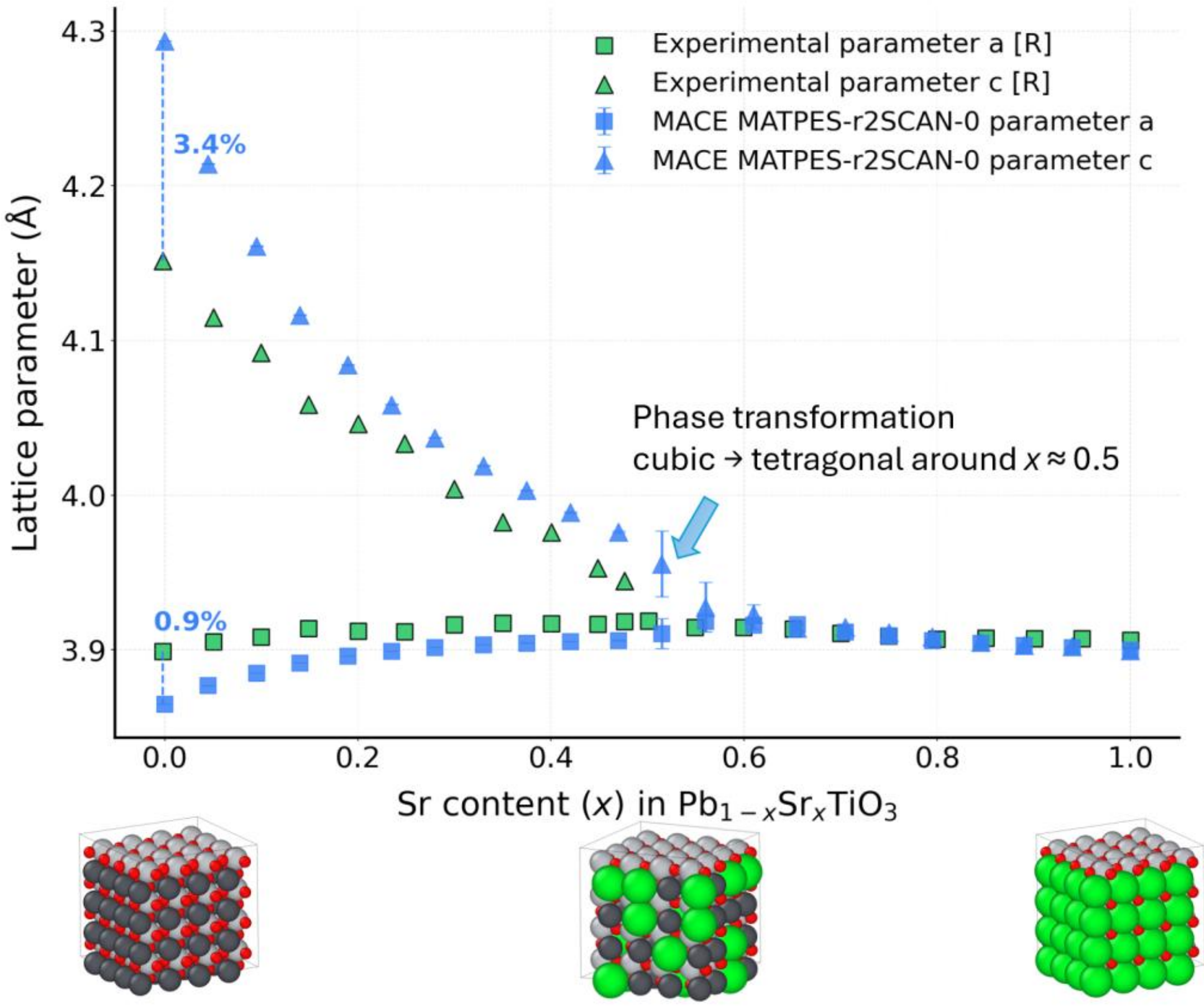

**Fig. 11:** The lattice parameters of $Pb_{1-x}Sr_xTiO_3$ as a function of Sr content calculated by universal MLIP MACE MATPES-r2SCAN from the generated SQSs, compared to the experimental reference ([R] = [42]).

The standard deviation across the five SQSs at each composition was negligible except near the transition region (x ≈ 0.5), where σ(a) = 0.01597 Å and σ(c) = 0.02114 Å, before decreasing again at other concentrations. Slight residual differences between calculated $a$ and $c$ in the cubic region were observed and are attributed to the finite size of the 320-atom supercell, but these are minor (less than 0.001 Å) and do not alter the compositional trends.

To assess the thermodynamic stability of the solid solution with respect to phase separation into $PbTiO_3$ and $SrTiO_3$ phases, we evaluated the composition-dependent mixing free energy $\Delta G_{\mathrm{mix}}(x,T) = \Delta E_{\mathrm{mix}}(x) - TS_{\mathrm{conf}}(x)$ of $Pb_{1-x}Sr_xTiO_3$, shown in **Fig. 12**. The entropy contribution is approximated as the configuration entropy. Therefore, the predicted temperature dependence reflects solely the effect of chemical disorder on the Pb-Sr sublattice site. At 0 K, the mixing energy is positive for compositions up to $x{\sim}0.6$ ($Pb_{0.4}Sr_{0.6}TiO_3$), indicating a thermodynamic driving force for phase separation, whereas for Sr-rich compositions ($x > 0.6$), it becomes negative, suggesting stabilization of the solid solution. Including configurational entropy lowers the free energy with increasing temperature. At 100 K, the mixing free energy remains positive over a reduced composition range( $x < 0.35$), while at 300 K it becomes negative across the entire composition range, indicating complete thermodynamic stabilization of the mixed phase. Error bars represent the standard deviation of energies obtained from five independently generated SQS configurations, demonstrating consistent trends and minor statistical variations.

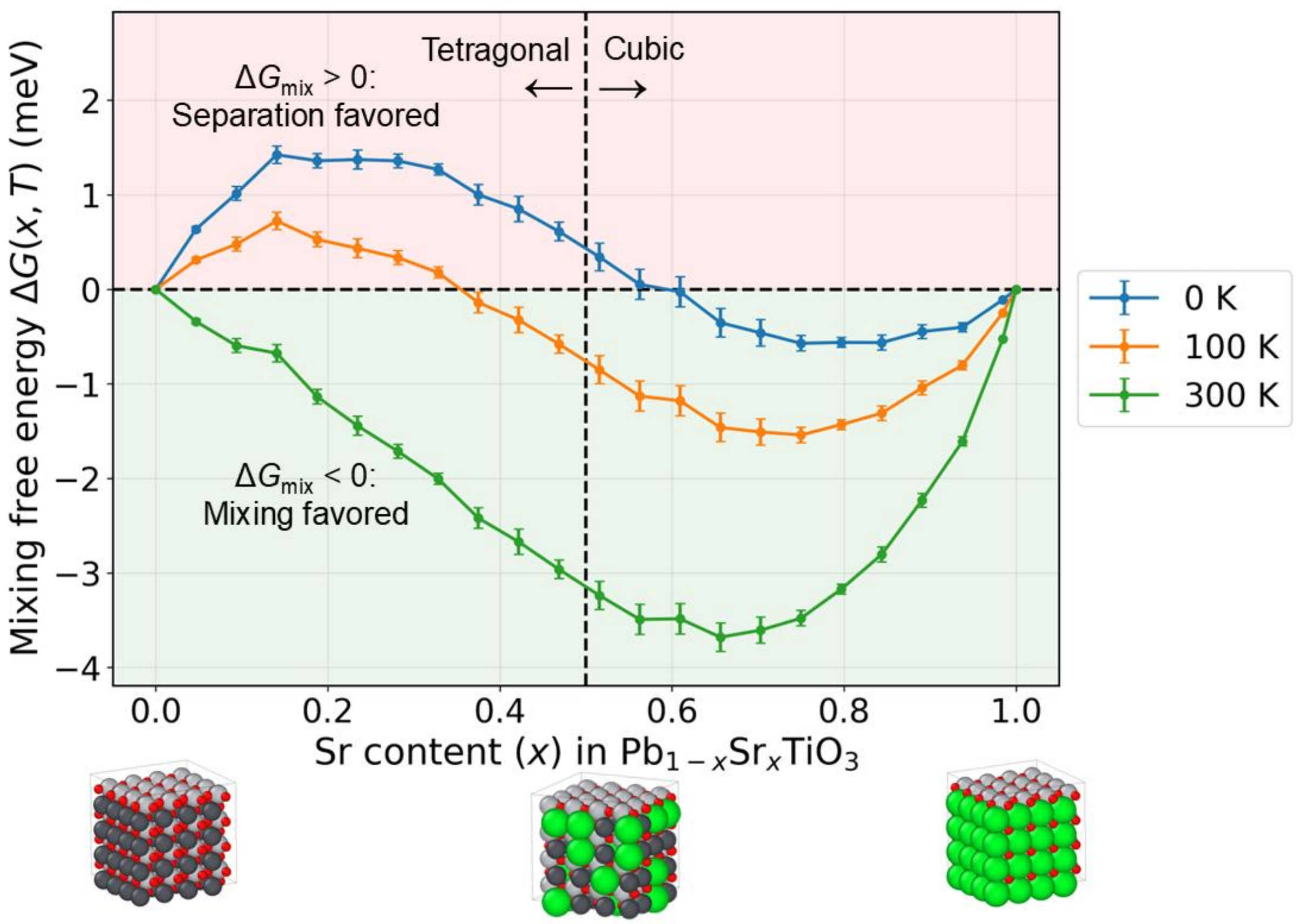

**Fig. 12:** Mixing energies (0 K) and mixing free energies (100 and 300 K) of $Pb_{1-x}Sr_xTiO_3$ as a function of Sr content ($x$), calculated per atom by universal MLIP MACE MATPES-r2SCAN from the relaxed SQSs. Finite-temperature curves include the ideal configurational entropy of Pb-Sr sublattice site.

## 5. Impact on User Adoption and Education

The presented interactive interface for preparing ATAT *mcsqs* input files significantly lowers the entry barrier for generating SQS with ATAT *mcsqs*, particularly for users with limited programming or command-line experience. Traditionally, the use of ATAT has required manual preparation of input files, familiarity with file formats, and often trial-and-error troubleshooting, all of which can discourage new users or students entering the field. By automating these steps through a graphical and intuitive interface, generating the necessary input files for SQS becomes far more straightforward.

From an adoption perspective, this simplification is expected to:

- Broaden the user base of ATAT by making it accessible to researchers outside computational materials science, such as experimentalists or educators.
- Reduce the dependency on community support forums by minimizing errors caused by file formatting or parameter misinterpretation in the input files.
- Make it easier to continue with simulations by allowing direct conversion of the best SQS into widely used formats, such as POSCAR (for VASP) or LAMMPS input files. This lets users move directly from SQS generation to DFT or MD simulations without extra file preparation.
- Improve the workflow for experienced users.

From an educational perspective, the interface also provides a valuable teaching tool. Instructors can integrate it into classroom demonstrations or computational labs to introduce students to the concept of random alloys and SQS generation without requiring prior coding skills. The guided workflows and integrated visualizations of structural data, correlation convergence, and radial distribution functions help students build intuition about the underlying theory.

Moreover, since *SimplySQS* is available online, it eliminates the need for local compilation and can be accessed from any device with an internet connection and a web browser.

## 6. Conclusions

In this work, we introduced *SimplySQS* (simplysqs.com), an automated and reproducible workflow for generating and analyzing special quasirandom structures using the ATAT *mcsqs* module. The online, interactive interface guides users through the complete process, from structure import and supercell setup to automated generation of ATAT input files, execution scripts, and post-processing of results. *SimplySQS* further enables direct visualization of *mcsqs* convergence, correlation values, and partial radial distribution functions, as well as conversion of optimized SQSs into common simulation formats. A key feature is the all-in-one bash script, which encapsulates the entire SQS search in a single, shareable file, ensuring reproducibility. In addition, *SimplySQS* provides a diagnostic tool to assess whether SQS optimization is recommended or whether random atomic occupation could be already sufficient for a given system size. By integrating these steps into a single platform, the usability of ATAT *mcsqs* is improved, reducing errors associated with manual file handling, and lowering the entry barrier for users with limited programming experience, while still benefiting advanced users through improved workflow efficiency and reproducibility. As a browser-based application, it requires no compilation and can be accessed from any device with an internet connection.

The usefulness of *SimplySQS* was illustrated on the perovskite series $Pb_{1-x}Sr_xTiO_3$ (PSTO, including $PbTiO_3$ and $SrTiO_3$), where a set of SQSs across the full composition range was generated using an all-in-one bash script. The resulting structures were subsequently optimized with the MACE MATPES-r$^2$SCAN-0 universal MLIP to determine the evolution of lattice parameters with increasing Sr content. The optimized PSTO structures accurately reproduced the experimentally observed cubic-to-tetragonal phase transformation near $x \approx 0.5$. In the cubic concentration region ($x > 0.5$), the simulated lattice parameters showed excellent agreement with experimental values, within a maximum of 0.3 % deviation. In the tetragonal region ($x < 0.5$), the lattice parameter *a* decreased slightly more (maximum deviation of 0.9 %), and parameter *c* increased more (maximum deviation of 3.4 %) than the experimental values, suggesting that the r$^2$SCAN-based universal MLIP may have insufficient underlying DFT training data for Pb-rich compositions. Beyond structural trends, we also evaluated mixing free energies from the relaxed SQS energies and the ideal configurational entropy of the Pb-Sr sublattice. At 0 K, the mixing energies are positive up to $x \sim 0.6$ (Pb-rich compositions), indicating a thermodynamic preference for phase separation into $PbTiO_3$ and $SrTiO_3$. However, the entropic contribution increasingly stabilizes the mixed phase with rising temperature, narrowing the composition range of positive mixing free energy at 100 K and yielding negative mixing free energies across the entire composition range at 300 K.

## 7. Acknowledgments

This work was supported by the Grant Agency of the Czech Technical University in Prague [grant No. SGS24/121/OHK2/3T/12].

## 8. Code Availability

The online version of *SimplySQS* is accessible at: https://simplysqs.com.

The underlying code is publicly available on the GitHub repository, containing documentation for the local installation: https://github.com/bracerino/atat-sqs-gui.

The all-in-one bash script generated by *SimplySQS* and used in the $Pb_{1-x}Sr_xTiO_3$ case study, together with the corresponding generated SQS structures, is available in the *Example* directory of the same repository.

Video tutorials demonstrating the use of the *SimplySQS* interface and the execution of the generated all-in-one scripts are available at: https://implant.fs.cvut.cz/atat-sqs-gui.